# Locally temporal-spatial pattern learning with graph attention mechanism for EEG-based emotion recognition

Yiwen Zhu [b], Kaiyu Gan [b], and Zhong Yin [a, b, *]

a. *Engineering Research Center of Optical Instrument and System, Ministry of Education, Shanghai Key Lab of Modern Optical System, University of Shanghai for Science and Technology, Shanghai, 200093, PR China*
b. *School of Optical-Electrical and Computer Engineering, University of Shanghai for Science and Technology, Shanghai, 200093, PR China*

**Abstract:** Technique of emotion recognition enables computers to classify human affective states into discrete categories. However, the emotion may fluctuate instead of maintaining a stable state even within a short time interval. There is also a difficulty to take the full use of the EEG spatial distribution due to its 3-D topology structure. To tackle the above issues, we proposed a locally temporal-spatial pattern learning graph attention network (LTS-GAT) in the present study. In the LTS-GAT, a divide-and-conquer scheme was used to examine local information on temporal and spatial dimensions of EEG patterns based on the graph attention mechanism. A dynamical domain discriminator was added to improve the robustness against inter-individual variations of the EEG statistics to learn robust EEG feature representations across different participants. We evaluated the LTS-GAT on two public datasets for affective computing studies under individual-dependent and independent paradigms. The effectiveness of LTS-GAT model was demonstrated when compared to other existing mainstream methods. Moreover, visualization methods were used to illustrate the relations of different brain regions and emotion recognition. Meanwhile, the weights of different time segments were also visualized to investigate emotion sparsity problems.

---

* Corresponding author: Zhong Yin, Tel.: +86 21 55271064. E-mail address: yinzhong@usst.edu.cn. Address: Jungong Road 516, Yangpu District, Shanghai 200093, P. R. China.

## 1. Introduction

Emotions are closely related to human cognitive functionality and play a fundamental role in regulating human behavior of perception and/or decision-making. In the field of human-computer interaction, technique of emotion recognition enables computers to automatically classify human affective states into discrete categories. Such an emotion recognizer facilitates building an intelligent module that makes machine agents adapt to human emotional fluctuations. When human users are collaborating with medical care devices [1], transportation systems [2], and/or industry robots [3], understanding accurate emotional states of human participants could improve overall efficiency of human-computer collaborations.

Emotion categories can be defined by discrete or dimensional models. The discrete model categorizes the affective state into six basic emotions, i.e., joy, sadness, surprise, fear, anger, and disgust. The dimensional model defines the continuous emotion based on core affective dimensions, e.g., valence and arousal scales. Emotions can be continuously labeled from positive to negative on the valence scale and passive to active on the arousal scale. As shown in Fig. 1, combinations of specific coordinates on these dimensions can elicit basic emotions defined by the discrete model.

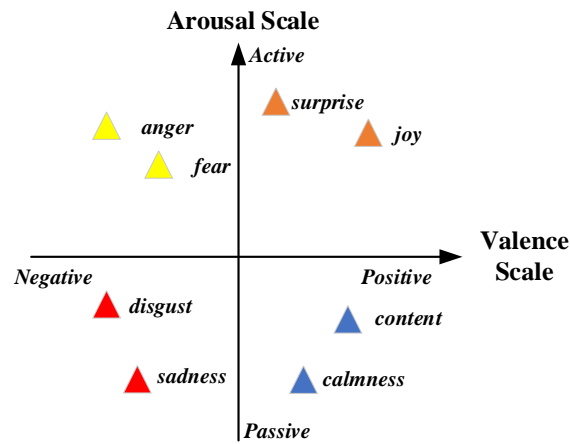

**Fig. 1**. Illustration of discrete and dimensional models for labeling emotions.

To measure emotional states, a variety of data modalities from human users have been employed, such as videos of body movements [4], images of facial expressions [5], records of speeches [6], and segments of psychophysiological signals [7]. Among these modalities, electroencephalography (EEG), a type of neurophysiological signal recorded from central nervous system, is particularly sensitive to identifying human inner cognitive states. The ongoing EEG is the sum of the voltage variations from pyramidal cells in different cortical regions. Applications of using the EEG to identify emotions have been reported in a series of studies [8,9]. The goal of this study is to build an EEG-based computational framework, i.e., an emotion recognizer, with the EEG as its input and emotion categories as its output.

Recent studies introduced advanced signal filtering and machine learning approaches to build EEG-based emotion recognizers [10,11]. The procedure mainly consists of four steps, i.e., EEG signal preprocessing, feature extraction, feature selection and/or feature learning, and training pattern classifiers. However, there are still three challenges required to be properly investigated. The first issue is that emotions may fluctuate in a wide range thus show a great sparsity and may lead to incorrect predictions. Second, the optimal spatial feature representation of the EEG is still unknown. It leads to difficulty in taking full use of the topological structure of the EEG energy shown by connections between different cortical regions and channels. The last issue is that the statistics and/or distributions of EEG features may drift greatly across different individuals. It always leads to a significant impairment in accuracy of individual-independent emotion recognizers.

Attempting to mitigate the above obstacles, we proposed a locally temporal-spatial pattern learning graph attention network, (termed LTS-GAT). Inspired by [12], we applied graph analysis to human cortical topology by considering each EEG channel as a node in the graph. Different from previous works [13,14], a graph attention network was employed instead of applying traditional spectral graph convolution. In particular, a divide-and-conquer scheme was used to examine local information on temporal and spatial dimensions of EEG patterns. It is aimed to quantitatively identify fluctuations of emotions over time instants and to discover spatial hidden variables according to the EEG channel paradigm. To improve the robustness against the inter-individual variations of the EEG statistics, a dynamical domain discriminator was added to the LTS-GAT to learn robust EEG feature representations across different participants.

In summary, the main contributions of our study are as follows:

1) We propose a novel LTS-GAT network to predict emotions based on the EEG. This model can capture local temporal information between different time instants and spatial information within and across brain regions.

2) A novel domain discriminator is proposed to alleviate distributional differences in the EEG data that are completely recorded from different individuals.

3) Importance of each brain region can be properly identified. The weights of specific attention modules can be used to interpret which cortical region plays the most significant role in measuring emotion variations.

4) The proposed emotion recognizer is validated via both individual-dependent and individual-independent classification tasks. Two publicly available EEG databases, HCI and DEAP, have been employed to evaluate the performance of the proposed algorithms.

The remainder of the paper is organized as follows. In Section 2, we review related works on EEG-based emotion recognition. Section 3 briefly summarizes preliminaries of graph theory and introduces the methods of applying graph analysis to the EEG. In Section 4, we elaborate on the feedforward structure of each computational module and the training algorithm of the LTS-GAT. Section 5 shows detailed experiments to validate generalization capability of the proposed method. Sections 6 and 7 provide some useful discussions and conclude the main findings.

## 2. Related Works

Selecting more representative and discriminative EEG features is an effective way to improve the accuracy of emotion recognition. Atkinson et al. [15] combined statistical-based feature selection methods to improve the performance of the individual-dependent emotion classification performance. A locally robust EEG feature selection method was proposed in [16] that could find invariant EEG indicators within a subset of individuals. A series of shallow learning models were applied to decode the EEG into emotion categories. In these works, support vector machine (SVM) [17] and $k$-nearest neighbors (kNN) [18] were widely used to identify emotions according to high-dimensional EEG features.

Recently, emotion recognition frameworks based on deep neural networks have attracted a lot of attention. Tzirakis et al. applied a deep convolutional neural network (ConvNet) as an emotion classifier. Cui et al. [19] proposed an end-to-end regional-asymmetric ConvNet that consists of temporal, regional, and asymmetric feature extractors to capture discriminative information. Although many previous studies have proved that the ConvNet shows great discriminant capability between different emotions, its feature learning pipeline is restricted to the 2-D Euclidean space. Considering that the EEG data are recorded via discrete 3-D locations in the spatial domain, only adopting the ConvNet structure may not be the optimal choice to take full use of the topological structure of cortices.

Graph neural network (GNN) was first proposed to deal with the data with a graph structure [20]. In early works, spectral convolution in the graph domain was extensively analyzed. Kipf et al. [21] proposed a standard graph convolutional neural network (GCN), which is a mature model combining GNN with spectral theory. Li et al. [22] proposed a curvature graph neural network, which exploits the structural properties of graph curvature effectively to make GNNs more adaptive to locating.

Considering the potential limitations of the classical GCN, non-spectral approaches have been investigated. It intuitively defines convolution operations on a spatial domain. Hechtlinger et al. [23] generalized the ConvNet to graph-structured data by utilizing a random walk procedure to unfold salient feature representations. Hamilton et al. [24] proposed GraphSAGE which is an inductive framework for generating node information on large graphs. This framework used a function that generated embedding by sampling and aggregating features from a local neighborhood around a node.

To improve the interpretability and capability of EEG feature learning with classical GCNs, we developed our model based on graph attention networks (GAT) [12]. On one hand, the EEG channels can be considered as nodes in the graph according to cortical topology which is more adaptive than using a 2-D matrix with ConvNets. Moreover, the GAT adopts the self-attention mechanism [25] to compute attention coefficients between each node and its neighbors. It thus can discover inner connections between different EEG channels and cortical regions related to emotion variation among multiple individuals.

## 3. Preliminary

### 3.1. Organizing the EEG Data with a Graph

Formally, a graph is defined as $\mathcal{G} = (\mathcal{V}, \mathcal{E})$, where $\mathcal{V}$ represents a set of the nodes and $\mathcal{E}$ denotes an edge set between nodes in $\mathcal{V}$. Each edge has a corresponding weight, which can be represented by a symmetric adjacency matrix $\mathbf{A} \in \mathbb{R}^{n \times n}$, where $n$ represents the node number and $a_{ij}$ denotes the edge weight between nodes $i$ and $j$. Then, a degree matrix $\mathbf{D} = diag(d_1, d_2, ..., d_n)$ is defined where each value on the diagonal is the sum of the corresponding row of the adjacency matrix $d_i = \sum_j a_{ij}$. When representing EEG data, each node is corresponding to an EEG channel. Thus, EEG features on the node set $\mathcal{V}$ can be represented by a matrix $\mathbf{X} \in \mathbb{R}^{n \times d}$, where $d$ denotes the dimension of feature on each node. After series of graph convolutional operations, the input $\mathbf{X}$ turns into output $\mathbf{Z} \in \mathbb{R}^{n \times d'}$, where $d'$ denotes the dimension of the learned EEG feature representation.

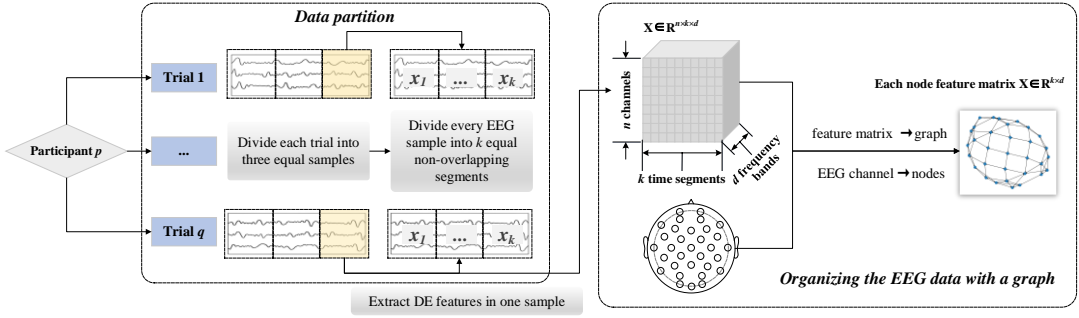

**Fig. 2.** Procedures for organizing the EEG data with a graph.

In the present study, the preprocessed EEG signals of one trial were divided into three samples with an equal length to achieve the data augmentation. To examine local temporal information, EEG signals in a sample were further divided into k non-overlapping segments with an equal length. Differential entropy (DE) features of all channels were extracted for each segment. Specifically, DE values within theta $\theta$ (4–7Hz), alpha $\alpha$ (8–12 Hz), beta $\beta$ (13–30 Hz), and gamma $\gamma$ (>30 Hz) bands were computed. To integrate graph analysis theory for representing the EEG data, the node number $n$ was equal to the number of the EEG channels. As a result, the feature tensor on the node set in one graph can be defined as $\mathbf{X} \in \mathbb{R}^{n \times k \times d}$, where $d$ denotes the number of frequency bands. The whole procedure for organizing the EEG data with a graph structure is shown in Fig. 2.

### 3.2. Graph Neural Network

There are two important operations in graph neural networks, i.e., aggregation and combination. For node $i$, an aggregating function $f_{AG}$ is defined to obtain information from its neighbor node $j$, i.e.,

$$a_i^{(l)} = f_{AG}^{(l)}(\{h_j^{(l-1)} : j \in \mathcal{N}_i\}).$$ (1)

In the equation, $\mathcal{N}_i$ represents the neighborhood of the node $i$. The terms $a_i$ and $h_j$ denote aggregation information and an implicit vector, respectively. The term $l$ denotes the index of the hidden layer. Classically, function $f_{AG}$ can be computed via spectral graph convolution [21] based on the symmetric adjacency matrix $\mathbf{A}$ and a degree matrix $\mathbf{D}$ with fixed weights as presented in Section 3.1 In this study, the $f_{AG}$ is defined via the graph attention mechanism to improve the capability for feature representation.

Then, a combining function $f_{CO}$ is introduced to fuse the information from the $l$-th hidden layer and implicit vectors of the $(l-1)$-th hidden layer,

$$h_i^{(l)} = f_{CO}^{(l)}(h_i^{(l-1)}, a_i^{(l)}).$$ (2)

Function $f_{CO}$ can be implemented with a concatenation operation combined with a fully connected layer. Moreover, a readout function $f_{RE}$ is used to obtain the final representation of the whole graph,

$$h_\mathcal{G} = f_{RE}(\{h_i | i \in \mathcal{G}\}). \tag{3}$$

Commonly, max pooling or average pooling functions are used as the $f_{RE}$ function. The $h_\mathcal{G}$ denotes learned feature representations with a graph structure.

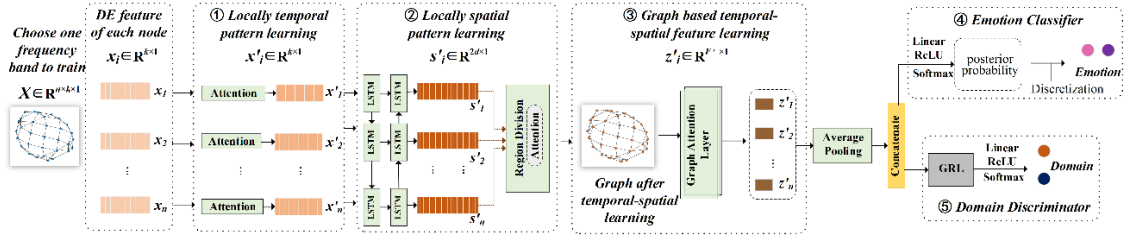

**Fig. 3.** Framework of the proposed LTS-GAT model for binary emotion recognition, where GRL denotes the gradient reversal layer.

## 4. Locally Temporal-spatial Pattern Learning Graph Attention Network

The proposed LTS-GAT framework can be summarized into four modules as shown in Fig. 3. First, a temporal pattern learning module is introduced to discover local information in the time domain across node features. Then, the derived feature matrix is fed into a bidirectional long short term memory (Bi-LSTM) network to learn high-level spatial feature abstractions. Specifically, local spatial structures of EEG feature distribution are represented by dividing electrodes into several subsets based on the local regions of the scalp. The learned feature matrix serves as the input of a GAT to predict emotion categories corresponding to the given EEG sample. For individual-independent tasks, a dynamical domain discriminator is implemented as an additional module to improve the capability of domain adaptation for the EEG data distribution between different individuals. The codes of the LTS-GAT model built in the study are available at [https://github.com/CFSRgroup/LTS-GAT-model](https://github.com/CFSRgroup/LTS-GAT-model).

### 4.1. Locally Temporal Pattern Learning

As described in Section 3, we could define $\mathbf{X} = [\mathbf{x}_1, \mathbf{x}_2, ..., \mathbf{x}_k] \in \mathbb{R}^{n \times k}$ as the extracted DE feature matrix for $k$ EEG segments in an EEG sample, where $\mathbf{x}_i$ denotes a feature vector extracted from the $i$-th segment and $n$ denotes the number of channels. It is noted that emotion intensity would fluctuate within a trial at different time instants [26]. Figure 4 shows a pattern of variations of the emotional intensity across consecutive EEG segments in an EEG sample. It may lead to the fact that the emotion labels obtained from the collected EEG data contain noise and are thus inconsistent with actual emotion categories.

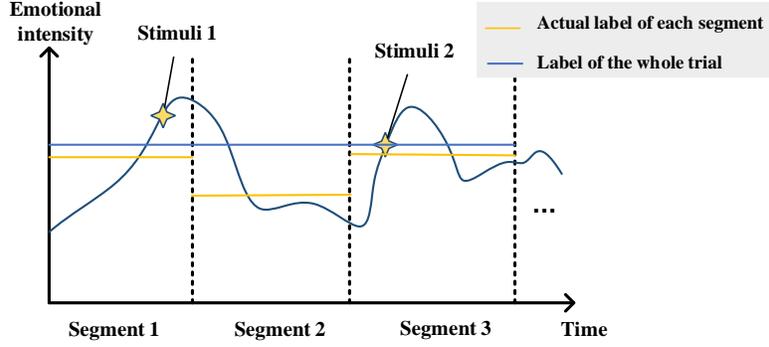

**Fig. 4.** Fluctuations of emotion intensity across different time segments in an EEG sample.

To alleviate the emotion sparsity existing in different segments, we applied an attention module to process feature time courses and compute segments' weights. For a DE feature matrix $\mathbf{X}$, three affine transformations are applied to obtain query, key and value matrices in an attention module [25]. The query matrix of temporal pattern learning module $\mathbf{Q}^{(t)} = \mathbf{W}_Q^{(t)T}\mathbf{X} + \mathbf{B}_Q^{(t)}$ with $\mathbf{Q}^{(t)} \in \mathbb{R}^{n \times k}$ is defined to compute the correlation between key matrix $\mathbf{K}^{(t)} = \mathbf{W}_K^{(t)T}\mathbf{X} + \mathbf{B}_K^{(t)}$ with $\mathbf{K}^{(t)} \in \mathbb{R}^{n \times k}$ and itself. Similarly, the value matrix is expressed as $\mathbf{V}^{(t)} = \mathbf{W}_V^{(t)T}\mathbf{X} + \mathbf{B}_V^{(t)}$ with $\mathbf{V}^{(t)} \in \mathbb{R}^{n \times k}$. Each column of $\mathbf{Q}^{(t)}$, $\mathbf{V}^{(t)}$, and $\mathbf{K}^{(t)}$ correspond to the EEG feature vector of a segment after the affine transformation.

To explore weights between different time segments, every key vector of $\mathbf{K}^{(t)}$ is supposed to compute attention coefficient with $\mathbf{Q}^{(t)}$. The process of attention mechanism is shown in Fig. 5 and the corresponding attention coefficient is defined as,

$$\mathbf{w}_i^{(t)} = \frac{\exp(\mathbf{k}_i^{(t)T}\mathbf{Q}^{(t)})}{\sum_{j=1}^{k}\exp(\mathbf{k}_j^{(t)T}\mathbf{Q}^{(t)})}. \tag{4}$$

In the equation, $i$ and $j$ represent different time segments, $\mathbf{k}_i^{(t)}$ is the $i$-th column vector of $\mathbf{K}^{(t)}$ that denotes the key vector of segment $i$. Here, the inner product operator is used as the scoring function for $\mathbf{k}_i^{(t)T}\mathbf{Q}^{(t)}$. Then, the weight matrix $\mathbf{W}_A^{(t)} = [\mathbf{w}_1^{(t)}, \mathbf{w}_2^{(t)}, ..., \mathbf{w}_k^{(t)}]^T \in \mathbb{R}^{k \times k}$ is defined to reveal importance of time segments quantitatively. Thus, the transformed feature matrix $\mathbf{X}' \in \mathbb{R}^{n \times k}$ containing local temporal information can be represented as,

$$\mathbf{X}' = \mathbf{V}^{(t)}\mathbf{W}_A^{(t)}. \tag{5}$$

Thus, the learnable parameter set of the locally temporal pattern learning module is $\mathbf{\theta}_A^{(t)} = \{\mathbf{W}_Q^{(t)}, \mathbf{B}_Q^{(t)}, \mathbf{W}_K^{(t)}, \mathbf{B}_K^{(t)}, \mathbf{W}_V^{(t)}, \mathbf{B}_V^{(t)}\}$.

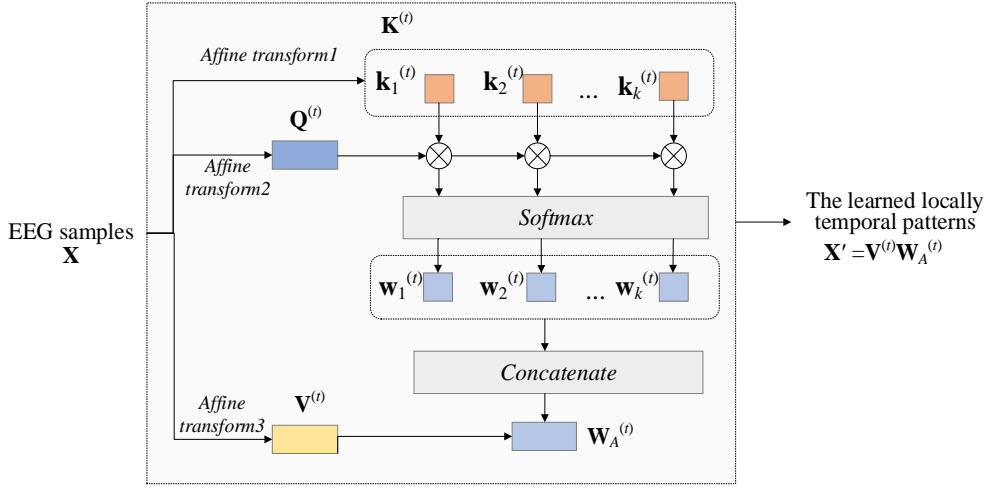

**Fig. 5.** Layout of the locally temporal pattern learning module.

### 4.2. Locally Spatial Pattern Learning

The spatial patterns of the EEG are learned with two stages. First, the feature matrix $\mathbf{X}'$ obtained by locally temporal pattern learning module is fed to a Bi-LSTM to discover the relationship between different EEG channels. Then, we divide EEG channels into several subsets according to their adjacent spatial locations and perform a region-attention operation.

Let $\mathbf{X}' = [\mathbf{x}_1', \mathbf{x}_2', ..., \mathbf{x}_n']^T \in \mathbb{R}^{n \times k}$ denotes the input of the Bi-LSTM and is used to learn spatial feature abstractions across EEG channels. We compare $n$ EEG channels with a continuous $n$-order sequence and each channel corresponds to one cell in the Bi-LSTM. In this case, for the $i$-th EEG cell, the input, forget, and output gates can be expressed as,

$$\boldsymbol{\mu}_i = \sigma(\mathbf{W}_{in} \cdot [\mathbf{h}_{i-1}^T, \mathbf{x}_i'^T]^T + \mathbf{b}_{in}),$$
$$\boldsymbol{\tau}_i = \sigma(\mathbf{W}_{fo} \cdot [\mathbf{h}_{i-1}^T, \mathbf{x}_i'^T]^T + \mathbf{b}_{fo}),$$
$$\mathbf{o}_i = \sigma(\mathbf{W}_{ou} \cdot [\mathbf{h}_{i-1}^T, \mathbf{x}_i'^T]^T + \mathbf{b}_{ou}).$$

(6)

In above equations, $\sigma(\cdot)$ denotes the sigmoid activation function and $\mathbf{x}_i'$ represents the input of the current cell. Matrices $\mathbf{W}_{in}$, $\mathbf{W}_{fo}$, and $\mathbf{W}_{ou}$ are the weights of input, forget, and output gates, respectively. The $\mathbf{h}_{i-1} \in \mathbb{R}^{2d \times 1}$ denotes a hidden state of the $(i-1)$-th cell with a hidden size of $d$. The terms, $\mathbf{b}_{in}$, $\mathbf{b}_{fo}$ and $\mathbf{b}_{ou}$, denote the corresponding bias of the three gates.

The state of the $i$-th cell $\tilde{\mathbf{c}}_i$ can be defined as,

$$\tilde{\mathbf{c}}_i = \tanh(\mathbf{W}_c \cdot [\mathbf{h}_{i-1}^T, \mathbf{x}_i^T]^T + \mathbf{b}_c).$$

(7)

In the equation, $\mathbf{W}_c$ denotes the weight matrix of the cell, $\mathbf{b}_c$ is the bias and $\mathbf{c}_{i-1}$ represents the state of $i$-th cell. Update the cell state $\mathbf{c}_i$ based on previous cell state and $\tilde{\mathbf{c}}_i$ with $\mathbf{c}_i = \boldsymbol{\tau}_i \circ \mathbf{c}_{i-1} + \boldsymbol{\mu}_i \circ \tilde{\mathbf{c}}_i$, where $\circ$ represents the Hadamard product. The output of a cell $\mathbf{h}_i$ is computed by $\mathbf{h}_i = \mathbf{o}_i \circ \tanh(\mathbf{c}_i)$ based on the cell state $\mathbf{c}_i$. Finally, for $n$ cells, feature matrix $\mathbf{X}'$ turns into $\mathbf{S} = [\mathbf{h}_1, \mathbf{h}_2, ..., \mathbf{h}_n]^T \in \mathbb{R}^{n \times 2d}$, which contains the learned location information among EEG channels. The learnable parameters are defined as $\boldsymbol{\theta}_L = \{\mathbf{W}_{in}, \mathbf{b}_{in}, \mathbf{W}_{fo}, \mathbf{b}_{fo}, \mathbf{W}_{ou}, \mathbf{b}_{ou}, \mathbf{W}_c, \mathbf{b}_c\}$.

Recent studies found that neuron responses to different emotion stimuli would be varied across different cortical areas [27]. Therefore, we divided feature map $\mathbf{S}$ into several local subsets. The region division paradigm based on 32 channels is illustrated in Fig. 6. Due to each cortical area containing different number of EEG channels, the dimension of local subsets varies. Thus, we employed affine transformations via $\{\mathbf{W}_g, \mathbf{B}_g\}$ to derive the identical dimension.

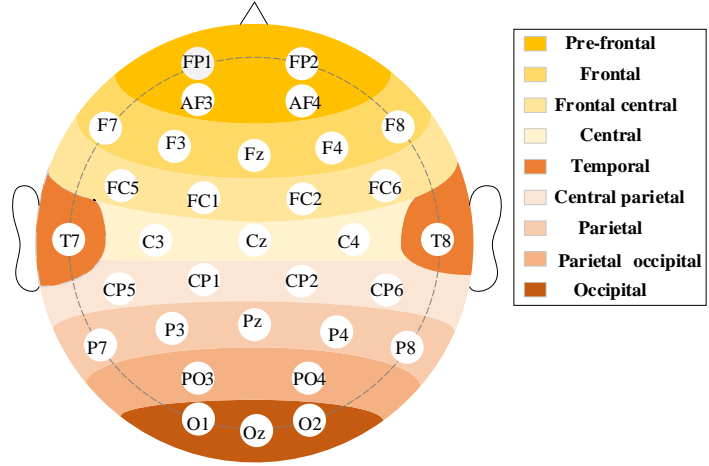

**Fig. 6.** Scheme of defining the EEG electrode subsets via different cortical regions.

Then, each subset corresponds to a cortical area which can be expressed as $\mathbf{G} = [\mathbf{g}_1, \mathbf{g}_2, ..., \mathbf{g}_N]^T \in \mathbb{R}^{N \times m}$ where $N$ represents the number of total regions and $m$ denotes the dimension of feature subset after the linear transformation. Next, we applied an attention layer to discover interactions of different regions. The attention coefficient of $i$-th region is calculated as,

$$\mathbf{w}_i^{(r)} = \frac{\exp(\mathbf{Q}^{(r)} \cdot \mathbf{k}_i^{(r)})}{\sum_{j=1}^{N} \exp(\mathbf{Q}^{(r)} \cdot \mathbf{k}_j^{(r)})}. \tag{8}$$

In the equation, $\mathbf{K}^{(r)} = [\mathbf{k}_1^{(r)}, \mathbf{k}_2^{(r)}, ..., \mathbf{k}_N^{(r)}] \in \mathbb{R}^{m \times N}$ and $\mathbf{Q}^{(r)} \in \mathbb{R}^{N \times m}$ denote the regional key and query matrices, respectively. The final output of spatial pattern learning module $\mathbf{H} \in \mathbb{R}^{N \times m}$ can be expressed as,

$$\mathbf{H} = \mathbf{W}^{(r)} \mathbf{V}^{(r)}, \tag{9}$$

where $\mathbf{W}^{(r)} = [\mathbf{w}_1^{(r)}, \mathbf{w}_2^{(r)}, ..., \mathbf{w}_N^{(r)}] \in \mathbb{R}^{N \times N}$ denotes the weight matrix and $\mathbf{V}^{(r)} \in \mathbb{R}^{N \times m}$ represents the regional value matrix. It is notable that the summation of each row of the $\mathbf{W}^{(r)}$ is used to illustrate the importance of each cortical area. The learnable parameters of the regional attention layer are defined as $\boldsymbol{\theta}_A^{(r)} = \{\mathbf{W}_g, \mathbf{B}_g, \mathbf{W}_Q^{(r)}, \mathbf{B}_Q^{(r)}, \mathbf{W}_K^{(r)}, \mathbf{B}_K^{(r)}, \mathbf{W}_V^{(r)}, \mathbf{B}_V^{(r)}\}$.

### 4.3. Graph based Temporal-Spatial Feature Learning

To take full use of the channel topology and the local temporal information from the learned $\mathbf{H}$, a GAT network was employed for high-level feature representation. Specifically, a graph attentional layer is used to learn graph-based geometrical information of the $\mathbf{H}$. For the output feature map $\mathbf{H} = [\mathbf{h}_1, \mathbf{h}_2, ..., \mathbf{h}_N]^T$ of spatial pattern learning module with $\mathbf{h}_i$ denoting the feature vector of the $i$-th region, affine transformations $\{\mathbf{W}_h, \mathbf{B}_h\}$ are applied on each $\mathbf{h}_i$ to achieve an identical dimension.

After that, the input of graph attentional layer is defined as $\mathbf{Z} = [\mathbf{z}_1, \mathbf{z}_2, ..., \mathbf{z}_n]^T \in \mathbb{R}^{n \times F}$, where $n$ denotes the number of nodes, $F$ denotes the number of features in one node, $\mathbf{z}_i$ is the feature set of node $i$. Then, a shared linear mapping is applied on $\mathbf{z}_i$ with a weight matrix of $\mathbf{W}_s \in \mathbb{R}^{F' \times F}$ with $F'$ representing the output feature dimension of each node. Next, a shared attentional mapping is employed [12]. The derived attention coefficient measures the importance of the EEG feature from the node $j$ to that from the node $i$. To embed into graph structure, assuming that $j \in \mathcal{N}_i$ with $\mathcal{N}_i$ denoting the set of neighbor nodes of node $i$ in the graph (including $i$), $j$ represents the first-order neighbor nodes of $i$ [12]. Then, compute attention coefficients for node $j$ with a softmax function,

$$a_{ij} = \frac{\exp(f_{Leaky}(\mathbf{a}_v^T[\mathbf{W}_s \mathbf{z}_i \,\|\, \mathbf{W}_s \mathbf{z}_j]))}{\sum_{k \in N_i} \exp(f_{Leaky}(\mathbf{a}_v^T[\mathbf{W}_s \mathbf{z}_i \,\|\, \mathbf{W}_s \mathbf{z}_k]))}, \tag{10}$$

where $\|$ denotes the concatenation operation, $\mathbf{a}_v^T$ is learnable weights, and $f_{Leaky}(\cdot)$ denotes the LeakyReLU activation function. Then, the output of node $i$ can be represented with multi-head attention [25],

$$\mathbf{z}_i' = f_{Leaky}(1/K \cdot \sum_{k=1}^{K} \sum_{j \in \mathcal{N}_i} a_{ij}^k \mathbf{W}_s^k \mathbf{z}_j). \tag{11}$$

Multi-head attention is employed to make the model obtain information jointly from different representation subspaces and reduce computational cost. We applied $K$ independent attention operators compute the coefficients in a parallel manner. Finally, the learned graph-based geometrical information is presented as $\mathbf{Z}' = \{\mathbf{z}_1', \mathbf{z}_2', ..., \mathbf{z}_n'\}, \mathbf{z}_i' \in \mathbb{R}^{F'}$. The learnable parameters in Eqns. (10)-(11) are defined as $\boldsymbol{\theta}_A^{(g)} = \{\mathbf{a}_v, \mathbf{W}_s\}$.

*4.4. Emotion Classifier with Domain Adaptation*

Predicting emotion categories corresponding to an EEG sample can be derived by the output of the graph attention network $\mathbf{Z}'$. To define the posterior probability of a specific emotion category, a softmax function $f_s(\cdot)$ is applied. For each input EEG sample $\mathbf{X}$, the posterior probability of binary emotion class categories $y \in \{0,1\}$ can be computed as,

$$p(\hat{y}|\mathbf{X}) = f_s(f_{pool}(\mathbf{Z}')). \tag{12}$$

In the equation, $f_{pool}(\cdot)$ adopts average pooling operations across all nodes on the graph to avoid deviation of the variance of the estimation. Learnable parameters of the emotion classifier are defined as $\mathbf{\theta}_c$. To train the classifier, the loss of a batch of training samples is defined as,

$$L_c(\mathbf{\theta}_p, \mathbf{\theta}_c \mid \mathbf{X}) = -\sum_{i=1}^{N_{batch}} L_{ce}(p(\hat{y}_i|\mathbf{X}_i), y_i), \tag{13}$$

where $\mathbf{\theta}_p = \{\mathbf{\theta}_A^{(s)}, \mathbf{\theta}_L, \mathbf{\theta}_A^{(r)}, \mathbf{\theta}_A^{(g)}\}$ are learnable parameters defined in previous sections, $N_{batch}$ represents the number of EEG samples in a batch, $L_{ce}(\cdot, \cdot)$ denotes the cross entropy loss function.

When the LTS-GAT is implemented with individual-independent paradigm, domain adaptation is adopted aiming to extract useful information from source domain (EEG data from a set of known individuals) and apply the transferable knowledge to the target domain (EEG data from an unknown individual). We assume that the source domain $\mathcal{D}_{so}$ consists of a set of labeled EEG samples $\mathbf{X}^{(so)}$ with the corresponding set of emotion categories $\mathbf{Y}^{(so)}$. For the target domain $\mathcal{D}_{ta}$, we have a set of unlabeled EEG samples $\mathbf{X}^{(ta)}$. A dynamical domain discriminator is introduced along with the emotion classifier to estimate the domain labels $\phi$, where 0 and 1 denote an EEG sample from the source and target domains, respectively.

Therefore, it is expected that the domain discriminator gradually unable to distinguish data from two domains along with the training process, which indicates the EEG distribution between $\mathcal{D}_s$ and $\mathcal{D}_t$ is similar to each other. Thus, the loss of domain discriminator should be maximized with a gradient reverse layer (GRL). The posterior probability of two domains can be calculated by,

$$p(\hat{\phi}|\mathbf{X}_i) = f_d(R_\lambda(\mathbf{Z}')), \tag{14}$$

where the feedforward operation of $R_\lambda(\cdot)$ is defined as $R_\lambda(x) = x$ and the feedback operation is defined as $dR_\lambda / dx = -\lambda I$ to reverse the gradient $I$. The term $\hat{\phi}$ denotes the estimated domain. It is noted that $\lambda$ is a parameter changing dynamically as iterations go on,

$$\lambda = \frac{2}{1 + \exp(-10 \cdot p)} - 1. \tag{15}$$

In the equation, $p$ represents the ratio of current iterations to total iterations.

By defining the learnable parameters of the domain adaptation module as $\boldsymbol{\theta}_d$, the overall loss function of the LTS-GAT can be expressed as,

$$L(\boldsymbol{\theta}_p, \boldsymbol{\theta}_c, \boldsymbol{\theta}_d \mid \mathbf{X}^{(so)}, \mathbf{X}^{(ta)}) = L_c(\boldsymbol{\theta}_p, \boldsymbol{\theta}_c \mid \mathbf{X}^{(so)}) - L_d(\boldsymbol{\theta}_p, \boldsymbol{\theta}_d \mid \mathbf{X}^{(so)}, \mathbf{X}^{(ta)}). \qquad (16)$$

In the equation, $L_d$ represents the cross-entropy loss of the domain discriminator. The feedforward and training procedure on EEG samples are summarized in Algorithm 1 and Algorithm 2 based on the ADAM optimizer [28].

**Algorithm 1.** Feedforward procedure of the LTS-GAT to learn the graph-based geometrical information

| | $\mathbf{Z}' = \textbf{LTS-GAT\_FF}\,(\,\mathbf{X}, \boldsymbol{\theta}_p\,)$ |
|---|---|
| **Input:** | An EEG sample $\mathbf{X}$ |
| | Parameters $\boldsymbol{\theta}_p$ of the LTS-GAT |
| **Output:** | Graph-based geometrical information $\mathbf{Z}'$ |
| 1 | Compute locally temporal attention mappings $\mathbf{Q}^{(t)}, \mathbf{K}^{(t)}, \mathbf{V}^{(t)}$ based on $\mathbf{X}$ |
| 2 | Compute $\mathbf{W}_A^{(t)}$ according to Eqn. (4) |
| 3 | $\mathbf{X}' \leftarrow \mathbf{V}^{(t)} \mathbf{W}_A^{(t)}$ |
| 5 | Compute the locally spatial pattern $\mathbf{S}$ with Eqns. (6)-(7) based on $\mathbf{X}'$ |
| 6 | Map $\mathbf{S} \rightarrow \mathbf{G}$ to match the region dimension |
| 7 | Compute regional attention mappings $\mathbf{Q}^{(r)}, \mathbf{K}^{(r)}, \mathbf{V}^{(r)}$ based on $\mathbf{G}$ |
| 8 | Compute $\mathbf{W}^{(r)}$ according to Eqn. (8) |
| 9 | $\mathbf{H} \leftarrow \mathbf{W}^{(r)} \mathbf{V}^{(r)}$ |
| 10 | Map $\mathbf{H} \rightarrow \mathbf{Z}$ to match the channel dimension |
| 11 | Compute graph-based geometrical mapping $\mathbf{Z}'$ using Eqns. (10)-(11) with $\mathbf{Z}$ |

**Algorithm 2.** Training procedure of the LTS-GAT.

| | $\boldsymbol{\theta}' = \textbf{LTS-GAT\_Train}(\,S_{ts}, S_{tr}^{(so)}, S_{tr}^{(ta)}, \boldsymbol{\theta}\,)$ |
|---|---|
| **Input:** | Numbers of batches $b$, epochs $n$, learning rate $\eta$ |
| | Training set $S_{tr} = \{\mathbf{X}_i\}_{i=1}^{b}$ |
| | Source domain set $S_{tr}^{(so)} = \{\mathbf{X}_i^{(so)}, y_i^{(so)}, \phi_i^{(so)}\}_{i=1}^{i=b}$ |
| | Target domain set $S_{tr}^{(ta)} = \{\mathbf{X}_i^{(ta)}, y_i^{(ta)}, \phi_i^{(ta)}\}_{i=1}^{i=b}$ |
| | Initialized parameters $\boldsymbol{\theta} = \{\boldsymbol{\theta}_p, \boldsymbol{\theta}_c, \boldsymbol{\theta}_d\}$ |
| **Output:** | Learned parameters $\boldsymbol{\theta}'$ |
| 1 | **for** $i = 1:n$, $l = 1:b$ |
| 2 | $p = (l + i \cdot b) / (n \cdot b)$ |
| 3 | Compute $\hat{\lambda}$ with Eqn. (15) |
| 4 | $\mathbf{Z}'_{i,l} \leftarrow \textbf{LTS-GAT\_FF}(\mathbf{X}_{i,l}, \boldsymbol{\theta}_p)$ |
| 5 | Compute $p(\hat{y}_{i,l}^{(so)} \mid \mathbf{X}_{i,l}^{(so)})$ with Eqn. (12) |
| 6 | Compute $p(\hat{\phi}_{i,l}^{(so)} \mid \mathbf{X}_{i,l}^{(so)}), p(\hat{\phi}_{i,l}^{(ta)} \mid \mathbf{X}_{i,l}^{(ta)})$ with Eqn. (14) |
| 7 | Compute $L_c(\boldsymbol{\theta}_p, \boldsymbol{\theta}_c \mid \mathbf{X}_{i,l}^{(so)})$ with Eqn. (13) |
| 8 | Compute $L_d(\boldsymbol{\theta}_p, \boldsymbol{\theta}_d \mid \mathbf{X}_{i,l}^{(so)}, \mathbf{X}_{i,l}^{(ta)})$ with Eqn. (13) |
| 9 | Compute $L(\boldsymbol{\theta}_p, \boldsymbol{\theta}_c, \boldsymbol{\theta}_d \mid \mathbf{X}_{i,l}^{(so)}, \mathbf{X}_{i,l}^{(ta)})$ with Eqn. (16) |
| 10 | $\nabla_{\boldsymbol{\theta}} L \leftarrow \left( (\nabla_{\boldsymbol{\theta}_p} L_c - \lambda \nabla_{\boldsymbol{\theta}_p} L_d), \nabla_{\boldsymbol{\theta}_c} L_c, \nabla_{\boldsymbol{\theta}_d} L_d \right)$ |
| 11 | Compute $\hat{\mathbf{m}}'$ and $\hat{\mathbf{v}}'$ based on $\nabla_{\boldsymbol{\theta}} L$ |
| 12 | $\boldsymbol{\theta} \leftarrow \boldsymbol{\theta} + \eta \cdot \hat{\mathbf{m}}' / (\hat{\mathbf{v}}'^{1/2} + \varepsilon)$ |
| 13 | $\boldsymbol{\theta}' \leftarrow \boldsymbol{\theta}$ |

Note: The $\varepsilon = 10^{-8}$ is set as the default value. Estimations of the first and second order gradients $\{\mathbf{m}', \mathbf{v}'\}$ are computed with the ADAM optimizer. The $\phi$ and $\hat{\phi}$ are the target and estimated domain labels, respectively. The $y$ and $\hat{y}$ are the target and estimated emotional labels, respectively.

## 5. Results

### 5.1. Database Descriptions

We validated the performance of the LTS-GAT on two public databases, HCI and DEAP. The HCI database recorded EEG data of 30 participants at a sampling frequency of 256 Hz with 32 active AgCl electrodes as shown in Fig. 6 [29]. Each participant watched 20 video clips (i.e., 20 trials) and each clip lasted between 34.9 and 117-s long. Each clip was selected to stimulate specific emotions such as happiness, sadness, disgust, and amusement. Similarly, subjective ratings with an interval from 1 to 9 on valence and arousal dimensions for each trial from 1 to 9 were collected. The data recorded from six participants were not analyzed due to the unfinished data collection.

The DEAP database is a multimodal dataset for emotion analysis which recorded EEG data of 32 participants [30]. Each participant was required to watch 40 one-minute-long clips of music videos (i.e., 40 trials). After participants watched each clip, they accomplished a self-assessment questionnaire by rating on valence, arousal, and dominance scales from a range of 1 to 9. All the EEG data are recorded at a sampling rate of 512 Hz using 32 active AgCl electrodes. The locations of these electrodes are identical to those employed in the HCI database.

### 5.2. EEG Data and Target Emotion Categories

For the two databases, all the EEG signals are downsampled to 128 Hz. Due to the motion and respiratory artifacts, a high pass filter with the cutoff frequency of 3 Hz and a bandpass filter with the cutoff frequencies of 4 and 45 Hz were applied to the HCI and DEAP databases, respectively. For the HCI database, we removed the 15-s baseline EEG and the 46-s EEG when the participants were performing subjective ratings in each trial. We then chose the last 30-s EEG of each trial for further analysis to avoid the label noise of emotion categories [31]. For the DEAP database, we removed the 3-second baseline signals and the length of filtered EEG of each trial is 60 s for each participant. For both databases, EEG signals from each trial were evenly divided into three samples. Then, DE features of all channels and four frequency bands were extracted from each sample. All the EEG features were standardized to zero mean and unit variance for each participant. We assigned two target emotion categories to each EEG sample based on the subjective ratings on arousal and valence dimensions for each trial. For arousal and valence ratings, the rating value higher than 5 indicates the high class and the value lower than 5 indicates the low class.

Hyper-parameter settings under two paradigms of the LTS-GAT are listed in Table 1.

Table 1. Hyper-parameter settings of the LTS-GAT under participant-dependent/independent classification paradigm.

| LTS-GAT model parameters | HCI Dependent /independent | DEAP Dependent /independent |
|---|---|---|
| Number of the EEG segments in a sample $k$ | 10/10 | 10/10 |
| Number of the cortical regions $N$ | 9/9 | 9/9 |
| Number of the nodes in a graph $n$ | 32/32 | 32/32 |
| Hidden size of Bi-LSTM | 16/48 | 32/32 |
| The number of graph attentional layer | 4/4 | 4/4 |
| Hidden size of GAT | 28/16 | 18/16 |
| Number of independent attention operators | 2/4 | 2/4 |
| Learning rate | 0.001/0.001 | 0.001/0.0001 |
| Batch size | 24/128 | 128/80 |
| Number of training epochs | 20/15 | 20/30 |

### 5.3. Individual-dependent Emotion Classification

For two databases, we both applied 10-fold cross validation on the EEG samples of each participant. That is, the training and testing EEG data were defined within a participant. In particular, the cross validation was performed at the video level. The training and testing EEG data were drawn from non-overlapped video clips. That is, to compare the performance, seven emotion recognizers were employed, i.e., support vector machine (SVM) [32], extreme gradient boosting (XGBoost) [33], ConvNet (LeNet-5) [34], Bi-LSTM [35], GCN [21], simple graph convolution (SGC) [36], and GAT [12].

Table 2. Participant-average accuracies/F1-score (%) of EEG-based emotion recognition methods in the participant-dependent classification paradigm.

| Method | HCI | | DEAP | |
|---|---|---|---|---|
| | Arousal | Valence | Arousal | Valence |
| SVM | 60.56/51.25 | 52.64/49.52 | 58.96/55.42 | 56.69/50.74 |
| XGBoost | 61.60/51.39 | 53.40/50.82 | 60.31/57.85 | 56.69/51.70 |
| CNN | 65.63/55.04 | 54.38/52.16 | 62.32/59.10 | 58.57/59.06 |
| BiLSTM | 63.54/55.82 | 54.79/52.56 | 62.08/57.27 | 61.17/61.48 |
| GCN | 61.53/53.16 | 52.08/48.82 | 61.28/58.03 | 56.46/57.18 |
| SGC | 62.85/54.76 | 54.38/54.37 | 61.25/56.53 | 56.51/58.49 |
| GAT | 60.69/52.28 | 52.92/48.67 | 60.03/56.52 | 55.57/55.33 |
| LTS-GAT | **65.76/59.25** | **57.71/55.83** | **63.39/59.99** | **66.74/68.07** |

Average classification performances over all participants on the HCI and DEAP databases are presented in Table 2. Inter-emotion discriminant capability of the arousal dimension is higher than that of the valence for the HCI database while the opposite observation is found for the DEAP. In general, the proposed model achieves higher F1-score and accuracy than other approaches on both databases. It is shown that the LTS-GAT has improved the average accuracy (or F1-score) by 3~11% (or 3~13%) compared to classical GAT model.

Table 3. Participant-average accuracies/F1-score (%) of EEG-based emotion recognition methods in the participant-independent classification paradigm.

| Method | HCI | | DEAP | |
|---|---|---|---|---|
| | Arousal | Valence | Arousal | Valence |
| SVM | 48.61/48.29 | 51.94/55.55 | 54.66/65.75 | 53.15/61.48 |
| RF | 50.42/48.22 | 48.82/55.22 | 56.90/68.23 | 54.04/64.43 |
| KPCA | 47.43/47.91 | 51.18/54.89 | 53.88/62.62 | 53.72/60.92 |
| TCA | 47.85/53.48 | 52.64/62.55 | 58.52/72.48 | 55.89/68.37 |
| ConvNet | 52.08/47.75 | 52.50/56.44 | 56.77/70.06 | 54.47/64.71 |
| t-LSTM | 51.74/54.26 | 51.11/52.86 | 57.50/70.74 | 55.03/64.61 |
| DANN | 53.40/52.53 | 52.22/51.76 | 58.91/**72.91** | 56.48/67.65 |
| BiDANN-R2 | 53.13/50.99 | 51.25/58.95 | 59.53/72.51 | 56.43/66.98 |
| LTS-GAT | **56.11/55.23** | 54.03/63.49 | **61.02**/72.53 | **66.17/69.11** |

*5.4. Individual-independent Emotion Classification*

The generalization capability of all emotion recognizers was also validated under the individual-independent paradigm for each database. We adopted leave-one-participant-out cross validation to calculate classifiers' performances. That is, the EEG data of one participant was used for testing and that of the other participants were taken as the training data to build the emotion recognizer. In every round of the cross validation, we defined training and testing data as the source and target domains to build the domain discriminator, respectively. To validate the performance, we adopted several transfer learning methods and state-of-the-art deep learning approaches for comparison purposes, which are listed as follows:

Table 4. Participant-average accuracies/F1-score (%) in the participant-dependent classification paradigm with different frequency bands from the HCI database.

| | Methods | $\theta$ | $\alpha$ | $\beta$ | $\gamma$ |
|---|---|---|---|---|---|
| | SVM | 58.40/51.67 | 62.15/54.86 | 61.25/54.89 | 57.99/48.87 |
| | XGBoost | 59.31/50.53 | 59.38/50.33 | 61.25/53.35 | 61.60/52.09 |
| | CNN | 62.99/54.96 | 64.31/55.24 | 63.54/54.84 | 63.26/52.96 |
| | Bi-LSTM | 63.68/52.83 | **65.76/55.96** | 64.03/54.51 | 62.85/53.85 |
| Arousal | GCN | 62.57/54.60 | 61.81/53.42 | 60.42/52.80 | 60.76/52.80 |
| | SGC | 62.71/54.02 | 62.78/54.80 | 62.99/54.88 | 62.78/55.48 |
| | GAT | 62.50/53.68 | 63.56/55.84 | 60.92/52.97 | 60.76/53.06 |
| | LTS-GAT | **64.58/57.10** | 64.17/55.96 | **64.79/57.21** | **64.97/56.94** |
| | | | | | |
| | SVM | 52.50/50.61 | 52.92/50.25 | 54.65/53.29 | 53.89/50.83 |
| | XGBoost | 51.39/50.15 | 51.46/50.57 | 55.35/54.58 | 54.10/52.21 |
| | CNN | 55.76/54.02 | 55.00/51.78 | 55.28/53.50 | 54.65/51.50 |
| | Bi-LSTM | 53.96/51.41 | 53.68/51.28 | 54.79/51.00 | 55.07/53.91 |
| Valence | GCN | 55.00/52.13 | 54.38/51.89 | 53.96/50.62 | 52.99/51.92 |
| | SGC | 54.17/52.78 | 54.17/53.46 | 55.63/53.09 | **55.38/52.05** |
| | GAT | 54.10/51.70 | 53.82/51.72 | 53.47/51.15 | 53.82/51.32 |
| | LTS-GAT | **56.39/54.14** | **56.67/54.12** | **56.88/55.26** | 55.14/52.83 |

(1) Two baseline methods: SVM [32] and random forest (RF) [37]. (2) Two transfer learning approaches: kernel principal component analysis (KPCA) [38] and transfer component analysis (TCA) [39]. (3) Four deep neural networks: ConvNet (with a LeNet-5 structure) [34], t-LSTM [35], DANN [40], and BiDANN-R2 [27].

Table 5. Participant-average accuracies/F1-score (%) in the participant-dependent classification paradigm with different frequency bands from the DEAP database.

| | Methods | $\theta$ | $\alpha$ | $\beta$ | $\gamma$ |
|---|---|---|---|---|---|
| | SVM | 56.67/53.92 | 54.48/51.60 | 56.95/53.32 | 57.50/54.14 |
| | XGBoost | 58.88/56.80 | 57.42/54.43 | 58.52/56.43 | 59.35/56.35 |
| | CNN | 62.03/59.49 | 61.88/58.63 | 61.30/57.43 | 62.63/60.07 |
| Arousal | Bi-LSTM | 61.02/57.42 | 60.94/57.68 | 61.17/56.58 | 62.45/57.73 |
| | GCN | 61.04/58.90 | 61.25/58.17 | 60.96/58.21 | 60.63/57.49 |
| | SGC | 61.85/57.71 | 61.43/56.96 | 61.09/56.38 | 61.80/57.41 |
| | GAT | 61.33/57.72 | 61.09/57.28 | 60.23/56.26 | 59.95/56.69 |
| | LTS-GAT | **63.15/60.78** | **62.97/61.15** | **62.99/60.82** | **63.02/60.76** |
| | | | | | |
| | SVM | 55.76/50.69 | 53.46/48.20 | 55.31/49.81 | 55.49/49.77 |
| | XGBoost | 54.84/51.52 | 54.11/50.68 | 54.37/50.80 | 54.79/50.37 |
| | CNN | 57.73/59.24 | 57.63/59.31 | 56.59/57.89 | 59.43/61.11 |
| Valence | Bi-LSTM | 60.18/61.07 | 58.59/59.32 | 59.17/59.88 | 60.78/61.52 |
| | GCN | 56.46/56.88 | 56.77/57.43 | 55.73/57.02 | 56.82/58.04 |
| | SGC | 57.42/59.24 | 56.17/58.78 | 55.89/57.96 | 57.55/60.11 |
| | GAT | 56.30/56.02 | 56.38/56.93 | 56.61/57.79 | 56.12/56.04 |
| | LTS-GAT | **66.77/67.55** | **66.46/67.77** | **66.98/68.08** | **66.17/67.45** |

To achieve a fair comparison, we used the same settings of the domain discriminator for defining the source and target domains for these approaches except for the SVM and RF models.

Table 3 shows the participant-average classification results of the HCI and DEAP databases. The LTS-GAT has better performance in the arousal dimension for the HCI database while better performance in the valence dimension for the DEAP. Overall, the proposed model achieved higher performance than other approaches on both databases except for the F1-score of the DANN of the arousal dimension for the DEAP. It is shown that the LTS-GAT improves the average accuracy (or F1-score) by 2~10% (or 2%~5%) compared to other advanced models with domain adaption.

Comparing with the results shown in Table 2, it can be concluded that the accuracy obtained in the participant-independent paradigm is worse than that obtained in the participant-dependent paradigm by around 0.6% to 9% for the LTS-GAT model due to discrepancies between EEG signals from different individuals.

*5.5. Individual-dependent Performance on Different Frequency Bands*

We further want to investigate the impacts of different frequency bands of EEG signals on individual-dependent emotion classification. Experiments were conducted on two databases using the DE features from four frequency bands, as reported in Tables 4 and 5. From the tables, the LTS-GAT achieved the highest accuracy from the theta and gamma band compared with all other methods on the HCI. It performs marginally worse than the Bi-LSTM and GCN in alpha and gamma bands. It also performs best for all four frequency bands on the DEAP database.

For both databases and emotional dimensions, performances of all emotion recognizers derived by the alpha, theta, and beta bands were insignificantly varied within 1%. One observation is found that the features from the gamma band impair the classification accuracy and F1-score in the valence scale of the HCI. Compared with Table 2, the classification performance of the LTS-GAT using all bands of DE features increased by 1~4% than using the single-band features.

### 5.6. Individual-independent Performance on Different Frequency Bands

Classification performance from four frequency bands on two datasets under the individual-independent paradigm is reported in Tables 6 and 7. It can be seen that the LTS-GAT achieves the highest accuracy in all frequency bands compared to most of the methods on two databases. The exception is F1-score of the DANN in the HCI database.

For both databases and emotional dimensions, performances of all emotion recognizers derived by four frequency bands are varied within 1~5%. Compared with Table 3, the classification performance of the LTS-GAT using all bands of DE features increased by 1~3% than adopting the single-band features.

Table 6. Participant-average accuracies/F1-score (%) in the participant-independent classification paradigm with different frequency bands from the HCI database

|  | Methods | $\theta$ | $\alpha$ | $\beta$ | $\gamma$ |
|---|---|---|---|---|---|
| *Arousal* | SVM | 50.83/50.61 | 51.25/50.85 | 49.03/48.73 | 48.19/48.44 |
|  | RF | 50.21/49.68 | 45.83/46.85 | 48.89/48.39 | 49.10/49.59 |
|  | KPCA | 51.11/51.05 | 49.65/49.37 | 49.86/49.02 | 48.89/49.09 |
|  | TCA | 52.36/50.55 | 47.36/45.37 | 47.57/45.74 | 49.86/48.00 |
|  | ConvNet | 51.11/47.50 | 50.56/46.51 | 51.86/48.09 | 50.49/48.66 |
|  | t-LSTM | 50.14/48.95 | 52.22/51.90 | 50.83/50.52 | 50.14/48.91 |
|  | DANN | 51.18/51.92 | 50.56/51.10 | 50.56/51.46 | 51.21/50.63 |
|  | BiDANN-R2 | 52.71/51.07 | 53.06/49.88 | 51.53/50.88 | 50.21/49.60 |
|  | LTS-GAT | **55.56/52.66** | **55.69/53.79** | **53.54/52.21** | **53.96/51.61** |
| *Valence* | SVM | 47.92/51.24 | 48.61/52.46 | 48.06/51.91 | 51.88/55.52 |
|  | RF | 52.36/59.90 | 50.97/58.46 | 50.56/57.19 | 51.67/59.04 |
|  | KPCA | 49.93/53.68 | 48.89/52.98 | 51.04/54.91 | 51.60/54.36 |
|  | TCA | 51.32/60.71 | 52.29/61.64 | 51.67/62.13 | 50.69/61.45 |
|  | ConvNet | 51.18/54.71 | 52.01/54.74 | 53.06/54.54 | 51.60/56.38 |
|  | t-LSTM | 51.11/54.94 | 52.78/57.00 | 51.18/57.44 | 51.04/57.07 |
|  | DANN | 51.04/59.24 | 51.53/60.70 | 53.26/63.08 | 51.67/60.12 |
|  | BiDANN-R2 | 52.29/59.16 | 50.14/56.30 | 52.85/**63.42** | 51.81/58.16 |
|  | LTS-GAT | **52.84/64.46** | **55.00/66.68** | **53.82**/61.64 | **53.47/65.06** |

Table 7. Participant-average accuracies/F1-score (%) in the participant-independent classification paradigm with different frequency bands from the DEAP database

| | Methods | $\theta$ | $\alpha$ | $\beta$ | $\gamma$ |
|---|---|---|---|---|---|
| *Arousal* | SVM | 54.40/64.44 | 54.79/65.33 | 53.65/65.05 | 56.33/67.39 |
| | RF | 55.96/69.11 | 55.86/68.68 | 54.92/67.49 | 55.34/67.70 |
| | KPCA | 54.61/63.99 | 54.27/63.84 | 54.77/64.08 | 54.82/64.12 |
| | TCA | 58.80/72.71 | 58.57/72.57 | 58.26/72.30 | 58.39/71.98 |
| | ConvNet | 57.47/71.02 | 56.74/70.58 | 57.55/71.30 | 58.12/71.64 |
| | t-LSTM | 57.92/70.86 | 56.85/70.65 | 57.29/70.72 | 57.14/70.83 |
| | DANN | 58.98/**72.95** | 58.88/**72.84** | 58.88/**72.88** | 58.78/72.73 |
| | BiDANN-R2 | 58.59/72.60 | 58.75/72.75 | 58.52/72.56 | 58.65/72.59 |
| | LTS-GAT | **60.16**/69.35 | **59.01**/66.39 | **59.22**/70.59 | **58.91/72.91** |
| | | | | | |
| *Valence* | SVM | 54.82/62.20 | 53.28/61.77 | 56.30/63.69 | 54.64/62.70 |
| | RF | 54.38/65.37 | 54.66/65.84 | 55.89/65.66 | 54.87/65.20 |
| | KPCA | 55.31/62.92 | 54.65/61.41 | 53.23/60.78 | 53.31/60.87 |
| | TCA | 55.73/68.18 | 55.29/66.72 | 57.71/69.21 | 55.73/68.18 |
| | ConvNet | 54.30/65.49 | 54.63/66.35 | 55.83/67.26 | 54.84/65.86 |
| | t-LSTM | 55.52/65.59 | 55.10/64.62 | 55.21/66.29 | 55.52/66.54 |
| | DANN | 56.71/67.54 | 56.28/**69.70** | 56.04/**69.11** | 56.48/68.77 |
| | BiDANN-R2 | 54.97/67.06 | 55.23/66.29 | 57.03/68.12 | 56.22/66.72 |
| | LTS-GAT | **66.41/68.96** | **66.02**/68.67 | **65.99**/68.72 | **66.15/69.07** |

## 6. Discussion

In this section, we first performed an ablation study to investigate contributions of each module for the LTS-GAT and then analyzed the importance of each time segment and cortical region for discriminating emotional states. The performance of the LTS-GAT is compared with the reported works in the literature.

### 6.1. Ablation Study

An ablation study was conducted to investigate the contribution of each module in the LTS-GAT. We first created several reduced versions of the network as follows. The LS-GAT denotes the locally temporal pattern learning module was removed and the LT-GAT represents the locally spatial pattern learning module was removed. We removed both the two modules to obtain a classical GAT model. When it comes to participant-independent classification tasks, we also tested a case where the domain discriminator was removed (denoted as "-DA"). Performances of reduced models are shown in Fig. 7. It is observed that the order of accuracies of four models is,

$$GAT < LT\text{-}GAT < LS\text{-}GAT < LTS\text{-}GAT,$$

in both arousal and valence dimensions in individual-dependent paradigm. While in individual-independent paradigm, there is an exceptional case where the GAT outperforms the LT-GAT.

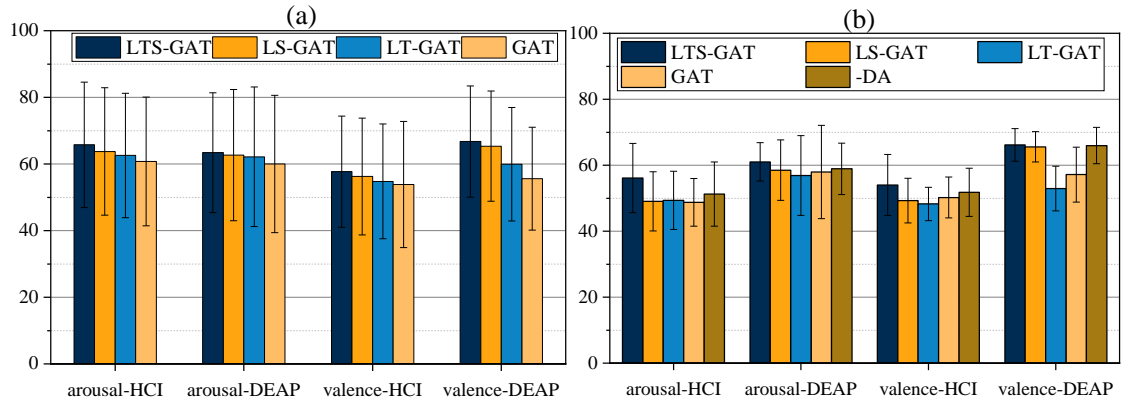

**Fig. 7.** Accuracies of the reduced models and the LTS-GAT: (a) Individual-dependent paradigm. (b) Individual-independent paradigm.

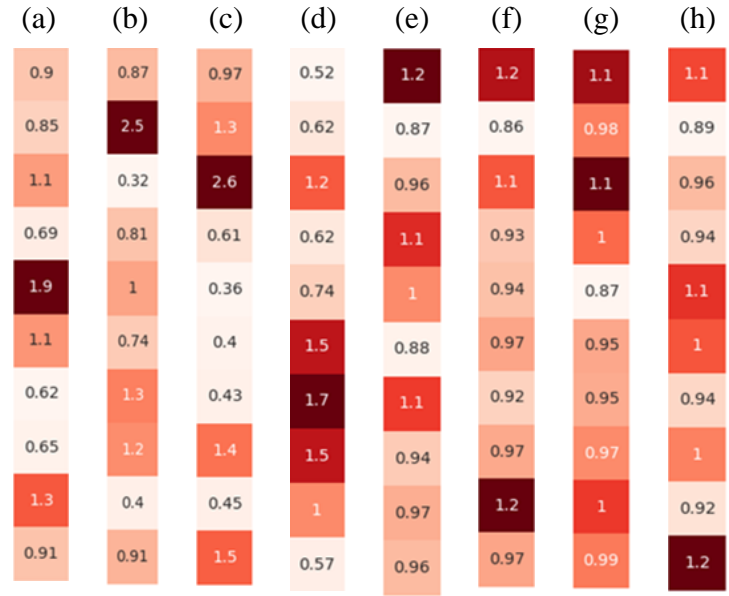

**Fig. 8.** Heatmaps based on the average weights of each time instant in an EEG sample derived from four different participants (a)-(d) and (e)-(h) from the HCI and DEAP database, respectively.

*6.2. Feature Importance Visualization*

We computed the average weights of each time instant defined in (1) from eight participants and drew heatmaps to investigate whether emotion sparsity exists in ten different time instants. As shown in Fig. 8, the range of the weight derived from the HCI participants is from 0.36 to 2.60, which is much larger than that of the DEAP (from 0.86 to 1.20). This fact can be the potential reason that the performance of the HCI of multiple classifiers is lower than that of the DEAP under the participant-dependent paradigm.

The importance of different cortical regions contributing to variations of affective states is shown in Fig. 9. We computed average attention weights defined in (11) of all participants in each database. Then, the weights are accurately mapped to cortical locations. We can observe that in the arousal dimension, the left pre-frontal and right-temporal regions are activated with higher weights for both databases and different bands. In the valence dimensions, frontal and parietal regions are highly activated for both databases and all frequency bands. These findings are consistent with the reported studies in [41,42].

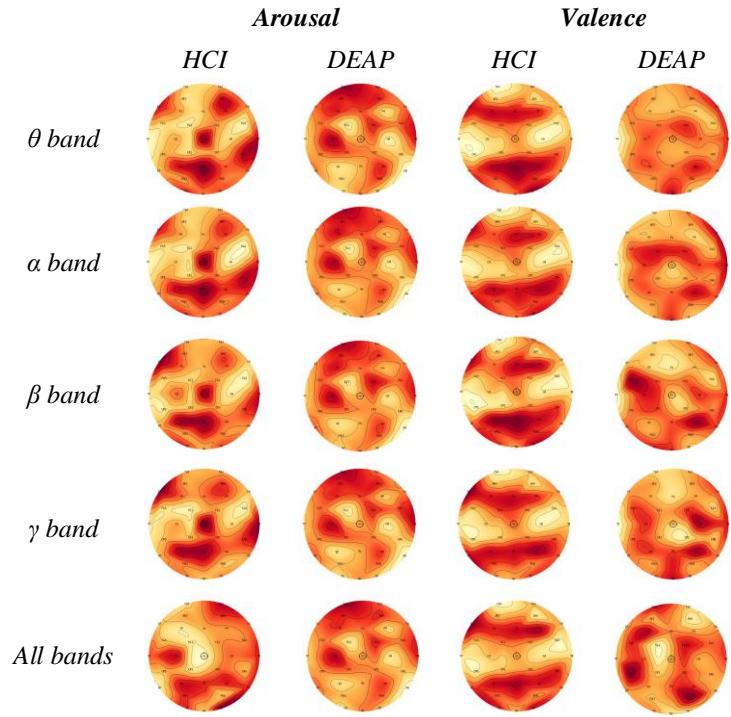

**Fig. 9**. Topographical maps based on the participant-average weights of cortical regions of different frequency bands.

### 6.3. Performance Comparison with Reported Studies

In Section 6.3, we enumerated recently reported results on two databases to evaluate our model furtherly. As shown in Table 8, on the DEAP database, the LTS-GAT obtains more discriminative patterns in the valence scale to improve the performance. The accuracy is improved by 8% using the LTS-GAT compared with [30,43] under the participant-dependent paradigm. In particular, our model improves the average accuracy by 6% for valence compared with [8], which is another advanced GNN-based method. On the HCI, it can be concluded that the LTS-GAT increases the accuracy by 0.5~3% and 3~7% for the arousal and valence scale under individual-dependent paradigms, respectively. However, it should be noted that the number of training samples, data processing methods, and definitions of the emotional classes could be different among these reported studies. These factors may induce lurking variables to compare the accuracy.

Table 8. Publication results on two databases of EEG-based emotion recognition

| Database | Methods | Arousal Acc/F1(%) | Valence Acc/F1(%) | Type |
|---|---|---|---|---|
| DEAP | Koelstra et al. [30] | 62.00/58.30 | 57.60/56.30 | D |
| DEAP | Li et al. [39] | 64.20/- | 58.40/- | D |
| DEAP | Zhuang et al. [44] | 71.99/- | 69.10/- | D |
| DEAP | Zhang et al. [5] | 65.39/- | 60.85/- | ID |
| DEAP | Song et al. [11] | 61.10/- | 59.29/- | ID |
| DEAP | Liang et al. [45] | 58.55/72.00 | 56.44/72.83 | ID |
| DEAP | Du et al. [46] | 72.97/- | 62.06/- | ID |
| **DEAP** | **LTS-GAT** | **63.39/59.99** | **66.74/68.07** | **D** |
| **DEAP** | **LTS-GAT** | **61.02/72.53** | **66.17/69.11** | **ID** |
| HCI | Gao et al. [47] | 63.00/62.80 | 56.90/54.70 | D |
| HCI | Momennezhad et al. [48] | 62.10/60.00 | 50.50/50.00 | D |
| HCI | Soleymani et al. [29] | 52.40/42.00 | 57.00/56.00 | ID |
| HCI | Liang et al. [45] | 62.06/62.05 | 60.64/72.18 | ID |
| HCI | Rayatdoost et al. [49] | 61.46/50.60 | 71.25/62.08 | ID |
| HCI | Huang et al. [50] | 61.80/- | 62.13/- | ID |
| HCI | Zhang et al. [51] | 65.20/63.41 | 65.37/65.34 | ID |
| **HCI** | **LTS-GAT** | **65.76/59.25** | **57.71/55.83** | **D** |
| **HCI** | **LTS-GAT** | **56.11/55.23** | **54.03/63.49** | **ID** |

Note: The term "D" denotes the individual-dependent paradigm and "ID" denotes the individual-independent paradigm.

## 7. Conclusion

In this paper, we proposed a locally temporal-spatial pattern learning graph attention network to facilitate EEG-based emotion classification. Our model was inspired by neuroscience findings that human emotion may be sparse in continuous time and affective responses can be closely related to the EEG entropy distribution between different channels and cortical sub-regions. We also introduced a dynamical domain discriminator to mitigate the distributional difference between EEG features across different participants. Our experimental results on two public datasets illustrate that the proposed LTS-GAT has better performance than several competitive methods under both individual-dependent and independent paradigms. Through analysis of cortical regions, we can infer that left pre-frontal, right-temporal, frontal, and parietal lobes are more important in EEG emotion classification. In terms of emotion sparsity, we can conclude that the HCI database possesses a potential higher sparsity than the DEAP database.

The main limitation of the study lies in two folds:

1) The classification performance under the individual-independent paradigm is far from perfect. Balancing the losses of the classifier and the discriminator may improve the generalization capability.

2) The proposed affective computing model only has the capability to distinguish emotions via a single modality, i.e., the EEG. Both the HCI and DEAP databases recorded physiological signals of the peripheral nervous system. Fusing information from multimodalities can potentially improve the performance of the emotion recognizer.

## CRediT Authorship Contribution Statement

**Yiwen Zhu**: Methodology, Investigation, Data processing, Code edition and debugging, Visualization, Writing- original draft, Writing- review & editing. **Kaiyu Gan**: Investigation, Code debugging, Visualization, Writing- review & editing. **Zhong Yin**: Methodology, Supervision, Writing- original draft, Writing- review & editing.

## Declaration of Competing Interest

The authors declare that they have no known competing financial interests or personal relationships that could have appeared to influence the work reported in this paper.

## Acknowledgements

This work is sponsored by the National Natural Science Foundation of China under Grant No. 61703277 and the Shanghai Sailing Program (17YF1427000).

## References

[1] A. Bayrakçi, E. Sert, N. Zorlu, A. Erol, A. Sariçiçek, L. Mete, Facial emotion recognition deficits in abstinent cannabis dependent patients, Compr. Psychiatry. 58 (2015) 160–164. https://doi.org/10.1016/j.comppsych.2014.11.008.

[2] G. Yang, Y. Lin, P. Bhattacharya, A driver fatigue recognition model based on information fusion and dynamic Bayesian network, Inf. Sci. (Ny). 180 (2010) 1942–1954. https://doi.org/10.1016/j.ins.2010.01.011.

[3] L. Chen, W. Su, Y. Feng, M. Wu, J. She, K. Hirota, Two-layer fuzzy multiple random forest for speech emotion recognition in human-robot interaction, Inf. Sci. (Ny). 509 (2020) 150–163. https://doi.org/10.1016/j.ins.2019.09.005.

[4] T. Zhu, Z. Xia, J. Dong, Q. Zhao, A Sociable Human-robot Interaction Scheme Based on Body Emotion Analysis, Int. J. Control. Autom. Syst. 17 (2019) 474–485. https://doi.org/10.1007/s12555-017-0423-5.

[5] F.Z. Canal, T.R. Müller, J.C. Matias, G.G. Scotton, A.R. de Sa Junior, E. Pozzebon, A.C. Sobieranski, A survey on facial emotion recognition techniques: A state-of-the-art literature review, Inf. Sci. (Ny). 582 (2022) 593–617. https://doi.org/10.1016/j.ins.2021.10.005.

[6] D. Li, Y. Zhou, Z. Wang, D. Gao, Exploiting the potentialities of features for speech emotion recognition, Inf. Sci. (Ny). 548 (2021) 328–343. https://doi.org/10.1016/j.ins.2020.09.047.

[7] X. Wu, B. Jiang, K. Yu, H. Chen, Separation and recovery Markov boundary discovery and its application in EEG-based emotion recognition, Inf. Sci. (Ny). 571 (2021) 262–278. https://doi.org/10.1016/j.ins.2021.04.071.

[8]    G. Zhang, M. Yu, Y.J. Liu, G. Zhao, D. Zhang, W. Zheng, SparseDGCNN: Recognizing Emotion from Multichannel EEG Signals, IEEE Trans. Affect. Comput. (2021). https://doi.org/10.1109/TAFFC.2021.3051332.

[9]    Z. Lan, O. Sourina, L. Wang, R. Scherer, G.R. Muller-Putz, Domain Adaptation Techniques for EEG-Based Emotion Recognition: A Comparative Study on Two Public Datasets, IEEE Trans. Cogn. Dev. Syst. 11 (2019) 85–94. https://doi.org/10.1109/TCDS.2018.2826840.

[10]   P.S. Ghare, A.N. Paithane, Human emotion recognition using non linear and non stationary EEG signal, in: Int. Conf. Autom. Control Dyn. Optim. Tech. ICACDOT 2016, Institute of Electrical and Electronics Engineers Inc., 2017: pp. 1013–1016. https://doi.org/10.1109/ICACDOT.2016.7877739.

[11]   Z. Lan, O. Sourina, L. Wang, Y. Liu, G.R. Muller-Putz, Stable Feature Selection for EEG-based Emotion Recognition, in: 2018 Int. Conf. Cyberworlds, 2018: pp. 176–183. https://doi.org/10.1109/CW.2018.00042.

[12]   P. Veličković, A. Casanova, P. Liò, G. Cucurull, A. Romero, Y. Bengio, Graph attention networks, 6th Int. Conf. Learn. Represent. ICLR 2018 - Conf. Track Proc. (2018) 1–12.

[13]   P. Zhong, D. Wang, C. Miao, EEG-Based Emotion Recognition Using Regularized Graph Neural Networks, IEEE Trans. Affect. Comput. (2020) 1–1. https://doi.org/10.1109/taffc.2020.2994159.

[14]   T. Song, W. Zheng, P. Song, Z. Cui, EEG Emotion Recognition Using Dynamical Graph Convolutional Neural Networks, IEEE Trans. Affect. Comput. 11 (2020) 532–541. https://doi.org/10.1109/TAFFC.2018.2817622.

[15]   J. Atkinson, D. Ca_Mpos, Improving BCI-based emotion recognition by combining EEG feature selection and kernel classifiers, Expert Syst. Appl. 47 (2016) 35–41. https://doi.org/10.1016/j.eswa.2015.10.049.

[16]   Z. Yin, W. Zhang, Z. Zheng, Locally Robust Feature Selection of EEG Signals for the Inter-subject Emotion Recognition, in: 2020 39th Chinese Control Conf., 2020: pp. 6250–6255. https://doi.org/10.23919/CCC50068.2020.9189239.

[17]   Y.P. Lin, C.H. Wang, T.P. Jung, T.L. Wu, S.K. Jeng, J.R. Duann, J.H. Chen, EEG-based emotion recognition in music listening, IEEE Trans. Biomed. Eng. 57 (2010) 1798–1806. https://doi.org/10.1109/TBME.2010.2048568.

[18]   R. Khosrowabadi, H.C. Quek, A. Wahab, K.K. Ang, EEG-based Emotion Recognition Using Self-Organizing Map for Boundary Detection, in: 2010 20th Int. Conf. Pattern Recognit., 2010: pp. 4242–4245. https://doi.org/10.1109/ICPR.2010.1031.

[19]   H. Cui, A. Liu, X. Zhang, X. Chen, K. Wang, X. Chen, EEG-based emotion recognition using an end-to-end regional-asymmetric convolutional neural network, Knowledge-Based Syst. 205 (2020) 106243. https://doi.org/10.1016/j.knosys.2020.106243.

[20]   F. Scarselli, M. Gori, A.C. Tsoi, M. Hagenbuchner, G. Monfardini, The graph neural network model, IEEE Trans. Neural Networks. 20 (2009) 61–80. https://doi.org/10.1109/TNN.2008.2005605.

[21]   T.N. Kipf, M. Welling, Semi-Supervised Classification with Graph Convolutional Networks, (2016). http://arxiv.org/abs/1609.02907.

[22]   H. Li, J. Cao, Z. Zhu, Y. Liu, Q. Zhu, G. Wu, Curvature graph neural network, Inf. Sci. (Ny). 592 (2022) 50–66. https://doi.org/10.1016/j.ins.2021.12.077.

[23]   Y. Hechtlinger, P. Chakravarti, J. Qin, A Generalization of Convolutional Neural Networks to Graph-Structured Data, (2017). http://arxiv.org/abs/1704.08165.

[24]   W.L. Hamilton, R. Ying, J. Leskovec, Inductive Representation Learning on Large Graphs, in: Proc. 31st Int. Conf. Neural Inf. Process. Syst., 2017: pp. 1025–1035. https://doi.org/10.48550/arXiv.1706.02216.

[25]   A. Vaswani, N. Shazeer, N. Parmar, J. Uszkoreit, L. Jones, A.N. Gomez, L. Kaiser, I. Polosukhin, Attention Is All You Need, (2017). http://arxiv.org/abs/1706.03762.

[26]   Y. Wang, Z. Huang, B. McCane, P. Neo, EmotioNet: A 3-D Convolutional Neural Network for EEG-based Emotion Recognition, in: 2018 Int. Jt. Conf. Neural Networks, 2018: pp. 1–7. https://doi.org/10.1109/IJCNN.2018.8489715.

[27]   Y. Li, W. Zheng, Y. Zong, Z. Cui, T. Zhang, X. Zhou, A Bi-Hemisphere Domain Adversarial Neural Network Model for EEG Emotion Recognition, IEEE Trans. Affect. Comput. 12 (2021) 494–504. https://doi.org/10.1109/TAFFC.2018.2885474.

[28]   D.P. Kingma, J. Ba, Adam: A Method for Stochastic Optimization, (2014). http://arxiv.org/abs/1412.6980.

[29]   M. Soleymani, J. Lichtenauer, T. Pun, M. Pantic, A multimodal database for affect recognition and implicit tagging, IEEE Trans. Affect. Comput. 3 (2012) 42–55. https://doi.org/10.1109/T-AFFC.2011.25.

[30]   S. Koelstra, C. Mühl, M. Soleymani, J.S. Lee, A. Yazdani, T. Ebrahimi, T. Pun, A. Nijholt, I. Patras, DEAP: A database for emotion analysis; Using physiological signals, IEEE Trans. Affect. Comput. 3 (2012) 18–31. https://doi.org/10.1109/T-AFFC.2011.15.

[31]   S. Katsigiannis, N. Ramzan, DREAMER: A Database for Emotion Recognition Through EEG and ECG Signals from Wireless Low-cost Off-the-Shelf Devices, IEEE J. Biomed. Heal. Informatics. 22 (2018) 98–107. https://doi.org/10.1109/JBHI.2017.2688239.

[32]   J. Suykens, L. Lukas, P. Van, D.B. De, M.J. Vandewalle, Least squares support vector machine classifiers: a large scale algorithm, Neural Process. Lett. 9 (1999) 293–300. https://doi.org/10.1023/A:1018628609742.

[33]   T. Chen, C. Guestrin, XGBoost: A Scalable Tree Boosting System, in: 22nd ACM SIGKDD Int. Conf. Knowl. Discov. Data Min., 2016: pp. 785–794. https://doi.org/https://doi.org/10.1145/2939672.2939785.

[34]    Y. LeCun, L. Bottou, Y. Bengio, P. Haffner, Gradient-based learning applied to document recognition, in: Proc. IEEE, 1998: pp. 2278–2323. https://doi.org/10.1109/5.726791.

[35]    Hochreiter, Sepp, Schmidhuber, Jurgen, Long short-term memory., Neural Comput. 9 (1997) 1735–1780. https://doi.org/https://doi.org/10.1162/neco.1997.9.8.1735.

[36]    F. Wu, T. Zhang, A.H. de Souza, C. Fifty, T. Yu, K.Q. Weinberger, Simplifying Graph Convolutional Networks, (2019). http://arxiv.org/abs/1902.07153.

[37]    L. Breiman, RANDOM FORESTS--RANDOM FEATURES, Mach. Learn. 45 (2001) 5–32. https://doi.org/doi:10.1023/A:1010933404324.

[38]    B. Schölkopf, A. Smola, K. Müller, Nonlinear Component Analysis as a Kernel Eigenvalue Problem, Neural Comput. 10 (1998) 1299–1319. https://doi.org/10.1162/089976698300017467.

[39]    S.J. Pan, I.W. Tsang, J.T. Kwok, Y. Qiang, Domain Adaptation via Transfer Component Analysis, IEEE Trans. Neural Networks. 22 (2011) 199–210. https://doi.org/10.1109/TNN.2010.2091281.

[40]    Y. Ganin, E. Ustinova, H. Ajakan, P. Germain, H. Larochelle, F. Laviolette, M. Marchand, V. Lempitsky, Domain-Adversarial Training of Neural Networks, J. Mach. Learn. Res. 17 (2016) 2096–2030. https://doi.org/10.1007/978-3-319-58347-1_10.

[41]    J.A. Coan, J.J.B. Allen, Frontal EEG asymmetry as a moderator and mediator of emotion, Biol. Psychol. 67 (2004) 7–50. https://doi.org/10.1016/j.biopsycho.2004.03.002.

[42]    A. Perry, S.N. Saunders, J. Stiso, C. Dewar, J. Lubell, T.R. Meling, A.K. Solbakk, T. Endestad, R.T. Knight, Effects of prefrontal cortex damage on emotion understanding: EEG and behavioural evidence, Brain. 140 (2017) 1086–1099. https://doi.org/10.1093/brain/awx031.

[43]    B. Li, Xiang; Zhang, Peng; Song, Dawei; Yu, Guangliang; Hou, Yuexian and Hu, EEG Based Emotion Identification Using Unsupervised Deep Feature Learning, in: SIGIR2015 Work. Neuro-Physiological Methods IR Res., 2015: pp. 1–2.

[44]    Z. Ning, Z. Ying, T. Li, Z. Chi, Z. Hanming, Y. Bin, Emotion Recognition from EEG Signals Using Multidimensional Information in EMD Domain, Biomed Res. Int. 2017 (2017) 8317357. https://doi.org/10.1155/2017/8317357.

[45]    Z. Liang, R. Zhou, L. Zhang, L. Li, G. Huang, Z. Zhang, S. Ishii, EEGFuseNet: Hybrid Unsupervised Deep Feature Characterization and Fusion for High-Dimensional EEG with an Application to Emotion Recognition, IEEE Trans. Neural Syst. Rehabil. Eng. 29 (2021) 1913–1925. https://doi.org/10.1109/TNSRE.2021.3111689.

[46]    X. Du, C. Ma, G. Zhang, J. Li, Y.K. Lai, G. Zhao, X. Deng, Y.J. Liu, H. Wang, An Efficient LSTM Network for Emotion Recognition from Multichannel EEG Signals, IEEE Trans. Affect. Comput. (2020). https://doi.org/10.1109/TAFFC.2020.3013711.

[47]    Z. Gao, S. Wang, Emotion Recognition from EEG Signals using Hierarchical Bayesian Network with Privileged Information, in: 5th ACM Int. Conf. Multimed. Retr., 2015: pp. 579–582. https://doi.org/10.1145/2671188.2749364.

[48]    Momennezhad, Ali, EEG-based emotion recognition utilizing wavelet coefficients, Multimed. Tools Appl. 77 (2018) 27089–27106. https://doi.org/10.1007/s11042-018-5906-8.

[49]    S. Rayatdoost, M. Soleymani, CROSS-CORPUS EEG-BASED EMOTION RECOGNITION, in: 2018 IEEE 28th Int. Work. Mach. Learn. Signal Process., 2018: pp. 1–6. https://doi.org/10.1109/MLSP.2018.8517037.

[50]    X. Huang, J. Kortelainen, G. Zhao, X. Li, A. Moilanen, T. Seppnen, M. Pietikinen, Multi-modal emotion analysis from facial expressions and electroencephalogram, Comput. Vis. Image Underst. 147 (2016) 114–124. https://doi.org/10.1016/j.cviu.2015.09.015.

[51]    W. Zhang, Z. Yin, Z. Sun, Y. Tian, Y. Wang, Selecting transferrable neurophysiological features for inter-individual emotion recognition via a shared-subspace feature elimination approach, Comput. Biol. Med. 123 (2020) 103875. https://doi.org/10.1016/j.compbiomed.2020.103875.